\documentclass[]{emulateapj}
\usepackage{graphicx}
\usepackage{array}
\usepackage{longtable}
\usepackage{natbib}
\usepackage{amssymb}
\begin{document}

\title{Pre-discovery and Follow-up Observations of the Nearby SN~2009nr: \\\
Implications for Prompt Type Ia SNe\altaffilmark{1}}
\author{Rubab~Khan\altaffilmark{2},
J.~L.~Prieto\altaffilmark{3,8},
G.~Pojma\'nski\altaffilmark{4},
K.~Z.~Stanek\altaffilmark{2,5},
J.~F.~Beacom\altaffilmark{2,5,6},
D.~M.~Szczygiel\altaffilmark{2},
B.~Pilecki\altaffilmark{4,7},
K.~Mogren\altaffilmark{2},
J.~D.~Eastman\altaffilmark{2},
P.~Martini\altaffilmark{2,5},
R.~Stoll\altaffilmark{2}
}

\altaffiltext{1}{Based on observations obtained using the 10-cm ASAS North telescope in Hawaii, the 50-cm Dedicated Monitor of Exotransit (DEMONEX) telescope at the Winer Observatory, the 2.4-m Hiltner telescope at the MDM Observatory, the 2.5-m du Pont telescope at the Las Campanas Observatory, the 3.5-m Astrophysical Research Consortium (ARC) telescope at the Apache Point Observatory, and the 6.5-m Magellan I (Baade) Telescope at the Las Campanas Observatory.}

\altaffiltext{2}{Dept.\ of Astronomy, The Ohio State University, 140
W.\ 18th Ave., Columbus, OH 43210; 
khan, kstanek, szczygiel, mogren, jdeast, martini, stoll@astronomy.ohio-state.edu}

\altaffiltext{3}{Carnegie Observatories, 813 Santa Barbara Street, 
Pasadena, CA 91101; jose@obs.carnegiescience.edu}

\altaffiltext{4}{Warsaw University Astronomical Observatory,
Al. Ujazdowskie 4, 00-478 Warsaw, Poland; gp, pilecki@astrouw.edu.pl}

\altaffiltext{5}{Center for Cosmology and AstroParticle Physics, 
The Ohio State University, 191 W.\ Woodruff Ave., Columbus, OH 43210}

\altaffiltext{6}{Dept.\ of Physics, The Ohio State University, 191 W.\
Woodruff Ave., Columbus, OH 43210; beacom@mps.ohio-state.edu}

\altaffiltext{7}{Dept.\ of Astronomy, University of Concepcion,
Casilla 160-C, Concepcion, Chile}

\altaffiltext{8}{Hubble, Carnegie-Princeton Fellow.}

\shorttitle{SN 2009nr and Prompt SN Ia}

\shortauthors{Khan et al.~2010}
 
\begin{abstract}
\label{sec:abstract}

We present photometric and spectroscopic observations of the Type Ia supernova SN~2009nr in UGC~8255 ($z=0.0122$). Following the discovery announcement at what turned out to be ten days after peak, we detected it at V $\simeq15.7$ mag in data collected by the All Sky Automated Survey (ASAS) North telescope 2 weeks prior to the peak, and then followed it up with telescopes ranging in aperture from 10-cm to 6.5-m. Using early photometric data available only from ASAS, we find that the SN is similar to the over-luminous Type Ia SN 1991T, with a peak at $M_V\simeq-19.6$ mag, and a slow decline rate of $\Delta m_{15}(B)\simeq0.95$~mag. The early post-maximum spectra closely resemble those of SN~1991T, while the late time spectra are more similar to those of normal Type Ia SNe. Interestingly, SN~2009nr has a projected distance of 13.0 kpc ($\sim4.3$ disk scale lengths) from the nucleus of the small star-forming host galaxy UGC~8255. This indicates that the progenitor of SN~2009nr is not associated with a young stellar population, calling into question the conventional association of luminous SNe~Ia with the ``prompt'' component directly correlated with current star formation. The pre-discovery observation of SN~2009nr using ASAS demonstrates the science utility of high cadence all sky surveys conducted using small telescopes for the discovery of nearby ($d\lesssim50$ Mpc) supernovae.
 
\end{abstract} 
\maketitle

\begin{table*}[ht]
\begin{center}
\caption{Optical Photometry of SN~2009nr}
\label{table:sn09nr_photometry}
\begin{tabular}{crccccccccc}
\hline 
\hline
\\
\multicolumn{1}{c}{HJD} &
\multicolumn{1}{c}{Phase} &
\multicolumn{1}{c}{B} &
\multicolumn{1}{c}{$\sigma$} &
\multicolumn{1}{c}{V} &
\multicolumn{1}{c}{$\sigma$} &
\multicolumn{1}{c}{R} &
\multicolumn{1}{c}{$\sigma$} &
\multicolumn{1}{c}{I} &
\multicolumn{1}{c}{$\sigma$} &
\multicolumn{1}{c}{Telescope /}
\\
\multicolumn{1}{c}{$-2450000$} &
\multicolumn{1}{c}{(days)} &
\multicolumn{1}{c}{(mag)} &
\multicolumn{1}{c}{} &
\multicolumn{1}{c}{(mag)} &
\multicolumn{1}{c}{} &
\multicolumn{1}{c}{(mag)} &
\multicolumn{1}{c}{} &
\multicolumn{1}{c}{(mag)} &
\multicolumn{1}{c}{} &
\multicolumn{1}{c}{Instrument}
\\
\hline
\hline
\\
5179.14 &  $-$14.0 & \dots & \dots & 14.51 & 0.06 & \dots & \dots & \dots & \dots & ASAS\\ 
5182.15 &  $-$11.0 & \dots & \dots & 14.51 & 0.06 & \dots & \dots & 14.89 & 0.16 & ASAS\\ 
5190.09 &  $-$3.1 & \dots & \dots & 13.86 & 0.07 & \dots & \dots & 14.12 & 0.19 & ASAS\\ 
5192.14 &  $-$1.0 & \dots & \dots & 13.77 & 0.06 & \dots & \dots & 14.05 & 0.16 & ASAS\\ 
5195.11 &  2.0 & \dots & \dots & 13.76 & 0.06 & \dots & \dots & 14.14 & 0.16 & ASAS\\ 
5198.09 &  4.9 & \dots & \dots & 13.70 & 0.06 & \dots & \dots & 14.37 & 0.17 & ASAS\\ 
5200.61 &  7.5 & \dots & \dots & 13.83 & 0.08 & \dots & \dots & 14.23 & 0.14 & ASAS\\ 
5206.09 &  12.9 & \dots & \dots & 14.10 & 0.08 & \dots & \dots & 14.57 & 0.19 & ASAS\\ 
5209.10 &  15.9 & \dots & \dots & 14.51 & 0.06 & \dots & \dots & \dots & \dots & ASAS\\ 
5214.13 &  21.0 & \dots & \dots & 14.51 & 0.06 & \dots & \dots & \dots & \dots & ASAS\\ 
5216.12 &  23.0 & \dots & \dots & 14.70 & 0.07 & \dots & \dots & 14.40 & 0.22 & ASAS\\ 
5218.09 &  24.9 & \dots & \dots & 14.82 & 0.07 & \dots & \dots & 14.68 & 0.24 & ASAS\\ 
5220.08 &  26.9 & \dots & \dots & 14.94 & 0.05 & \dots & \dots & 14.53 & 0.21 & ASAS\\ 
5222.09 &  28.9 & \dots & \dots & 15.03 & 0.06 & \dots & \dots & 14.70 & 0.23 & ASAS\\ 
5225.56 &  32.4 & \dots & \dots & 15.27 & 0.08 & \dots & \dots & 14.72 & 0.29 & ASAS\\ 
5232.12 &  39.0 & \dots & \dots & \dots & \dots & \dots & \dots & 15.09 & 0.20 & ASAS\\ 
5239.05 &  45.9 & \dots & \dots & 14.51 & 0.06 & \dots & \dots & \dots & \dots & ASAS\\ 
5253.83 &  60.7 & \dots & \dots & 16.34 & 0.04 & 16.03 & 0.03 & 16.13 & 0.02 & DEMONEX\\ 
5268.88 &  75.7 & \dots & \dots & 16.66 & 0.04 & 16.50 & 0.03 & \dots & \dots & DEMONEX\\ 
5271.25 &  78.1 & \dots & \dots & 16.78 & 0.03 & 16.60 & 0.05 & 16.86 & 0.04 & DEMONEX\\ 
5271.78 &  78.6 & \dots & \dots & 16.72 & 0.03 & 16.55 & 0.02 & \dots & \dots & DEMONEX\\ 
5276.69 &  83.5 & \dots & \dots & 16.83 & 0.05 & 16.81 & 0.05 & \dots & \dots & DEMONEX\\ 
5276.69 &  83.5 & 17.41 & 0.05 & 16.74 & 0.02 & 16.62 & 0.03 & \dots & \dots & du Pont-WFCCD\\ 
5277.75 &  84.6 & 17.39 & 0.05 & 16.80 & 0.02 & 16.77 & 0.02 & \dots & \dots & du Pont-WFCCD\\ 
5278.80 &  85.6 & 17.40 & 0.05 & 16.83 & 0.01 & 16.80 & 0.01 & \dots & \dots & du Pont-WFCCD\\ 
5279.76 &  86.6 & 17.41 & 0.05 & 16.84 & 0.03 & 16.84 & 0.01 & \dots & \dots & du Pont-WFCCD\\ 
5280.90 &  87.7 & \dots & \dots & 17.10 & 0.04 & 17.01 & 0.05 & \dots & \dots & DEMONEX\\ 
5296.99 &  103.8 & \dots & \dots & 17.45 & 0.07 & 17.50 & 0.06 & \dots & \dots & DEMONEX\\ 
5303.82 &  110.7 & 17.79 & 0.04 & 17.43 & 0.01 & 17.67 & 0.01 & \dots & \dots & MDM-OSMOS\\ 
\hline
\hline
\end{tabular}
\end{center}
\end{table*}

\begin{table*}[!ht]
\begin{center}
\caption{Journal of Spectroscopic Observations of SN~2009nr}
\label{table:sn09nr_spectroscopy}
\begin{tabular}{crccccc}
\hline 
\hline
\\
\multicolumn{1}{c}{HJD} &
\multicolumn{1}{c}{Phase} &
\multicolumn{1}{c}{Spectral Range} &
\multicolumn{1}{c}{Resolution} &
\multicolumn{1}{c}{Slitwidth} &
\multicolumn{1}{c}{Exptime} &
\multicolumn{1}{c}{Telescope /}
\\
\multicolumn{1}{c}{($-2450000$)} &
\multicolumn{1}{c}{(days)} &
\multicolumn{1}{c}{(\AA)} &
\multicolumn{1}{c}{(\AA)} &
\multicolumn{1}{c}{(\arcsec)} & 
\multicolumn{1}{c}{(s)} &
\multicolumn{1}{c}{Instrument}
\\
\hline
\hline
\\
5205.0 & 12 & 3850-7510 & 15 & 2.0 & 600 & MDM-CCDS  \\ 
5212.0 & 19 & 3850-7510 & 15 & 2.0 & 600 & MDM-CCDS  \\ 
5247.0 & 54 & 4350-8020 & 15 & 2.0 & 1200 & MDM-CCDS \\ 
5253.0 & 60 & 3515-7180 & 15 & 2.0 & 1200 & MDM-CCDS \\ 
5275.8 & 83 & 3500-9500 & 4 & 0.9 & 3600 & Magellan-IMACS \\
5277.7 & 85 & 3700-9300 & 7 & 1.7 & 1200 & du Pont-WFCCD \\ 
5278.8 & 86 & 3700-9300 & 7 & 1.7 & 1800 & du Pont-WFCCD \\ 
5279.7 & 87 & 3700-9300 & 7 & 1.7 & 1200 & du Pont-WFCCD \\
5295.9 & 103 & 3500-9600 & 7 & 1.5 & 1800 & APO-DIS
\\
\hline
\hline
\end{tabular}
\end{center}
\end{table*}

\section{Introduction}
\label{sec:introduction}

The absence of hydrogen and the abundant presence of Si~II close to maximum light distinguish Type Ia supernovae (SNe~Ia) from other types of SNe~\citep[e.g.,][]{ref:Filippenko_1997}. They are astrophysically significant for several reasons. SNe~Ia are reliable cosmological distance indicators because their intrinsic luminosity can be well calibrated, and were used to discover the acceleration of the universe and to characterize dark energy models~\citep[e.g.,][]{ref:Riess_1998,ref:Perlmutter_1999,ref:Astier_2006,ref:Wood-Vasey_2007,ref:Kessler_2009}. The heavy elements (mostly Fe group) produced by their explosions play a pivotal role in the chemical evolution of galaxies~\citep[e.g.,][]{ref:Kobayashi_2009}. Well observed SNe~Ia let us investigate the nature of their progenitors and study their explosion mechanism.

The nature of the progenitors of SNe~Ia is still being debated. In the single degenerate model, a carbon/oxygen white dwarf (CO WD) in a close binary system accretes matter from a stellar companion~\citep{ref:Whelan_1973}. As its mass approaches or exceeds the Chandrasekhar limit, it is unable to support itself through electron degeneracy pressure. The resulting thermonuclear explosions are expected to release similar amounts of energy, as they all burn the same amount of fuel ignited through the same mechanism~\citep[e.g.,][]{ref:Hillenbrandt_2000}. In the double-degenerate model, SN~Ia results from the merger of two degenerate compact objects, igniting the potentially  super-Chandrasekhar mass progenitor~\citep[e.g.,][]{ref:Iben_1984,ref:Webbink_1984}. Recent studies have argued that the double-degenerate scenario is the dominant mechanism~\citep[e.g.,][]{ref:Pritchet_2008,ref:Ruiter_2009,ref:Mennekens_2010}.

Early discovery of SNe enables us to study their photometric evolution, and identify spectroscopic features that are difficult to detect in post-maximum spectra dominated by Fe group elements. Moreover, well sampled photometric data around the peak are crucial for the light curve fits needed to calibrate the luminosity. These provide valuable information regarding the nature of the progenitor and the explosion itself. The All-Sky Automated Survey for variable stars (ASAS) has two small telescopes with wide-field cameras, at sites in Chile and Hawaii, and is obtaining images of the entire sky every $\sim3$ days down to V $\simeq15$ magnitude~\citep{ref:ASAS_1997,ref:ASAS_2002,ref:ASAS_2010}. The high cadence, large area, and magnitude limit of this survey are optimal for making early discoveries of bright SNe in nearby galaxies.

\begin{figure*}[ht]
\begin{center}
\plotone{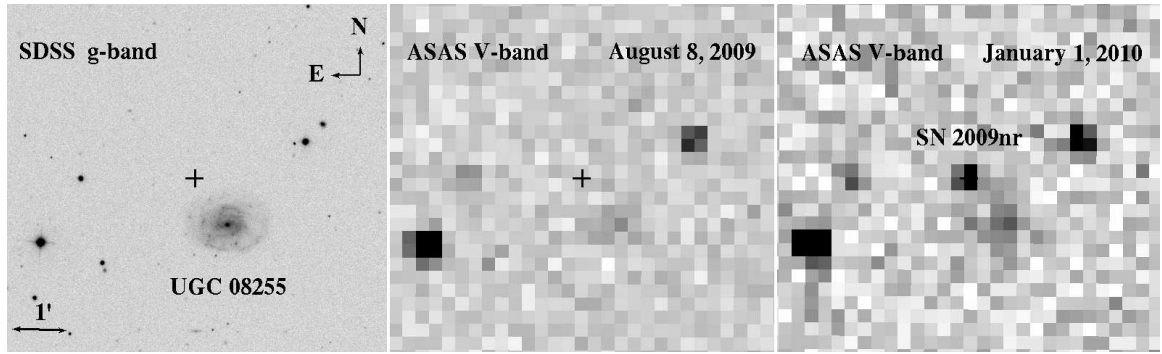}
\end{center}
\caption{SDSS g-band image of the SN~2009nr host galaxy UGC~8255
  (left) and ASAS V-band images of the same field taken prior to
  (center) and after (right) the explosion. Each image has the same
  orientation and shows the same region. The location of SN~2009nr is
  marked with a cross~\citep[R.A.=13$^h$10$^m$58\fs95,
    Dec.=+11\arcdeg29\arcmin29\farcs3; J2000.0;][]{ref:CBET_2111}.}
\label{fig:host_sdss}
\end{figure*}

SN~2009nr (R.A.=13$^h$10$^m$58\fs95, Dec.=+11\arcdeg29\arcmin29\farcs3; J2000.0), in the Scd galaxy UGC~8255 ($z=0.0122$), was detected in an unfiltered image taken at Blagoveschensk, Russia on 2009 Dec.~22.901 UT, and the discovery was announced on 2010 Jan.~6 UT by the Master Robotic Telescope Network~\citep{ref:CBET_2111}. It was also independently found by the Lick Observatory Supernova Survey (LOSS) in an unfiltered image taken with the 0.76-m Katzman Automated Imaging Telescope (KAIT) on 2010 Jan.~6.53~\citep{ref:CBET_2095}. It was spectroscopically confirmed as a 1991T-like Type Ia by the CfA Supernova Program using spectra taken on 2010 Jan.~7 UT using the Whipple Observatory 1.5-m telescope~\citep{ref:CBET_2112}. Following the discovery announcement, we detected the SN in archival data collected by the ASAS North telescope on 2009 Dec.~13 UT, and then followed it up with telescopes ranging in aperture from 10-cm to 6.5-m. Figure~\ref{fig:host_sdss} shows the SDSS g-band image of the SN~2009nr host galaxy UGC~8255, and ASAS V-band images of the same field taken prior to and after the explosion.

Our analysis shows that the earliest ASAS observation of SN~2009nr was obtained 14 days prior to maximum light ($2455193.16\pm0.3$; 2009 Dec.~27 UT), and 24 days prior to the discovery announcement. Our spectroscopic and photometric follow-up observations of the SN and its host galaxy lasted until 110 days after the peak. In what follows, Section~\ref{sec:data} describes the photometric and spectroscopic observations, data reduction, and calibration methods. Section~\ref{sec:analysis} presents the optical light curves, spectral evolution, and host galaxy properties of SN~2009nr. Section~\ref{sec:discussion} discusses the implications of our observations for conventional understandings of the nature of prompt Type Ia SNe, and the utility of ASAS for studying nearby SNe. Section~\ref{sec:conclusion} presents our conclusions.

\section{Observations and Data Reduction}
\label{sec:data}

\subsection{Photometric Data}
\label{sec:phot}

The photometric data were collected using the 10-cm ASAS North
telescope in Hawaii~\citep{ref:ASAS_2002,ref:Pigulski_2009}, the 50-cm
Dedicated Monitor of Exotransit (DEMONEX) telescope located at the
Winer Observatory~\citep{ref:Eastman_2009}, the 2.4-m Hiltner
telescope at the MDM Observatory, and the 2.5-m du Pont telescope at
the Las Campanas Observatory. Table \ref{table:sn09nr_photometry} has
the calibrated $BVRI$ magnitudes of SN~2009nr from the telescopes and
instruments we used.

The data reduction and photometry from the ASAS North telescope images
were obtained through a custom-made, well-tested data reduction
pipeline described in detail in \cite{ref:ASAS_1998,ref:ASAS_2002}. 
The $V$ and $I$ magnitudes form ASAS were tied to the Johnson~$V$ and 
Cousins~$I$ scales using Tycho~\citep{ref:Hog_2000} and 
Landolt~\citep{ref:Landolt_1983} stars.

The images from DEMONEX, the Ohio State Multi-Object Spectrograph~\citep[OSMOS;][]{ref:Stoll_2010,ref:Martini_2010} on the MDM 2.4-m telescope, and the Wide-Field CCD (WFCCD) on
the du Pont telescope were reduced (bias subtraction, cosmic ray (CR) rejection,
and flat-fielding) using standard techniques in IRAF\footnote{IRAF is
  distributed by the National Optical Astronomy Observatories, which
  are operated by the Association of Universities for Research in
  Astronomy, Inc., under cooperative agreement with the National
  Science Foundation.}. Relative photometry of the supernova with
respect to several local standards ($\geq 4$) was obtained with PHOT
in IRAF using an aperture comparable to the FWHM of the stars in the
image, and a background estimated in an annulus around the object. Since the SN is
well isolated and far from the host galaxy, aperture photometry was
the technique of choice. We did not apply any fringing correction to 
the late-time $I$-band images (both from DEMONEX). 

We calibrated the relative photometry in the $BVRI$ system using
zero-point offsets estimated with respect to the magnitudes of the
local standards from the SDSS-DR7 release \citep{ref:DR7_2009},
which were transformed from the SDSS filter system to the standard
Bessell/Kron-Cousins magnitudes using the transformations in the
SDSS-DR7 website\footnote{\scriptsize{\tt
www.sdss.org/dr7/algorithms/sdssUBVRITransform.html\#Lupton2005}}. The
uncertainties quoted in Table \ref{table:sn09nr_photometry} include
the standard errors and the rms of the zero-point transformation from
the local standards.

\subsection{Spectroscopic Data}
\label{sec:spec}

We obtained 9 post-maximum optical spectra of SN~2009nr using the
Boller and Chivens CCD Spectrograph (CCDS), WFCCD on the 2.5-m du Pont
telescope at the Las Campanas Observatory, the Dual Imaging
Spectrograph (DIS) on the 3.5-m Astrophysical Research Consortium
(ARC) telescope at the Apache Point Observatory, and the
Inamori-Magellan Areal Camera \& Spectrograph (IMACS; \cite{ref:Dressler_2006}) 
on the 6.5-m Magellan I (Baade) telescope at Las Campanas
Observatory. Table \ref{table:sn09nr_spectroscopy} shows the journal
of spectroscopic observations of SN~2009nr, with details of the epochs
and different spectrographs used.

The spectroscopic data were reduced using standard techniques in IRAF,
which included the basic data reductions (bias subtraction, CR
rejection, and flat-fielding), 1D spectrum extraction, wavelength
calibration with HeNeAr (WFCCD, DIS, and IMACS) and Xe (CCDS)
arc-lamps obtained at the same position as observations, and flux
calibration using a spectroscopic standard observed the same night as
the science observations. The spectra from CCDS were obtained using a
N-S oriented slit, while the spectra from DIS and WFCCD were obtained
with the slit oriented at the parallactic angle. The IMACS spectrum
was obtained with a ${\rm PA} = 55.4 \arcdeg$ to align the slit along
the SN and the center of the host galaxy, in order to obtain spectra
of H~II regions in the host. We did not attempt to do absolute
spectrophotometry, and the spectrophotometric calibration is only
relative.

\section{Analysis}
\label{sec:analysis}

\begin{figure}[!ht]
\begin{center}
\plotone{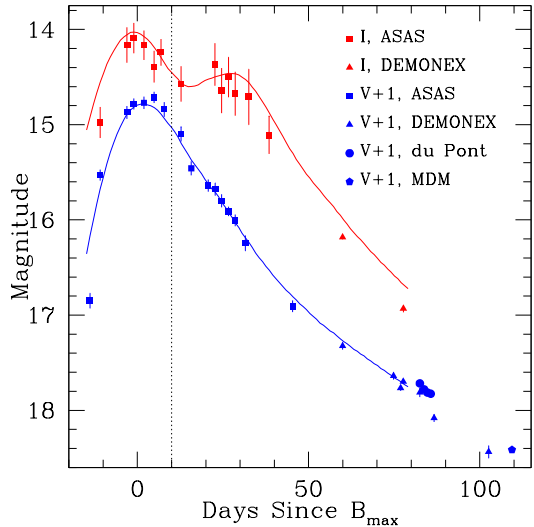}
\end{center}
\caption{V and I band light curves of SN~2009nr. The
  horizontal axis is in days from the date of maximum
  brightness in the B-band. The two solid lines are SN~Ia light curve
  template fits to the ASAS data points only. The dotted line shows
  the epoch of the discovery announcement ($\sim10$ days after
  $B_{max}$). The V-band light curve is shifted by +1 magnitude
  for clarity.}
\label{fig:lightcurve}
\end{figure}

\begin{figure}[!ht]
\begin{center}
\plotone{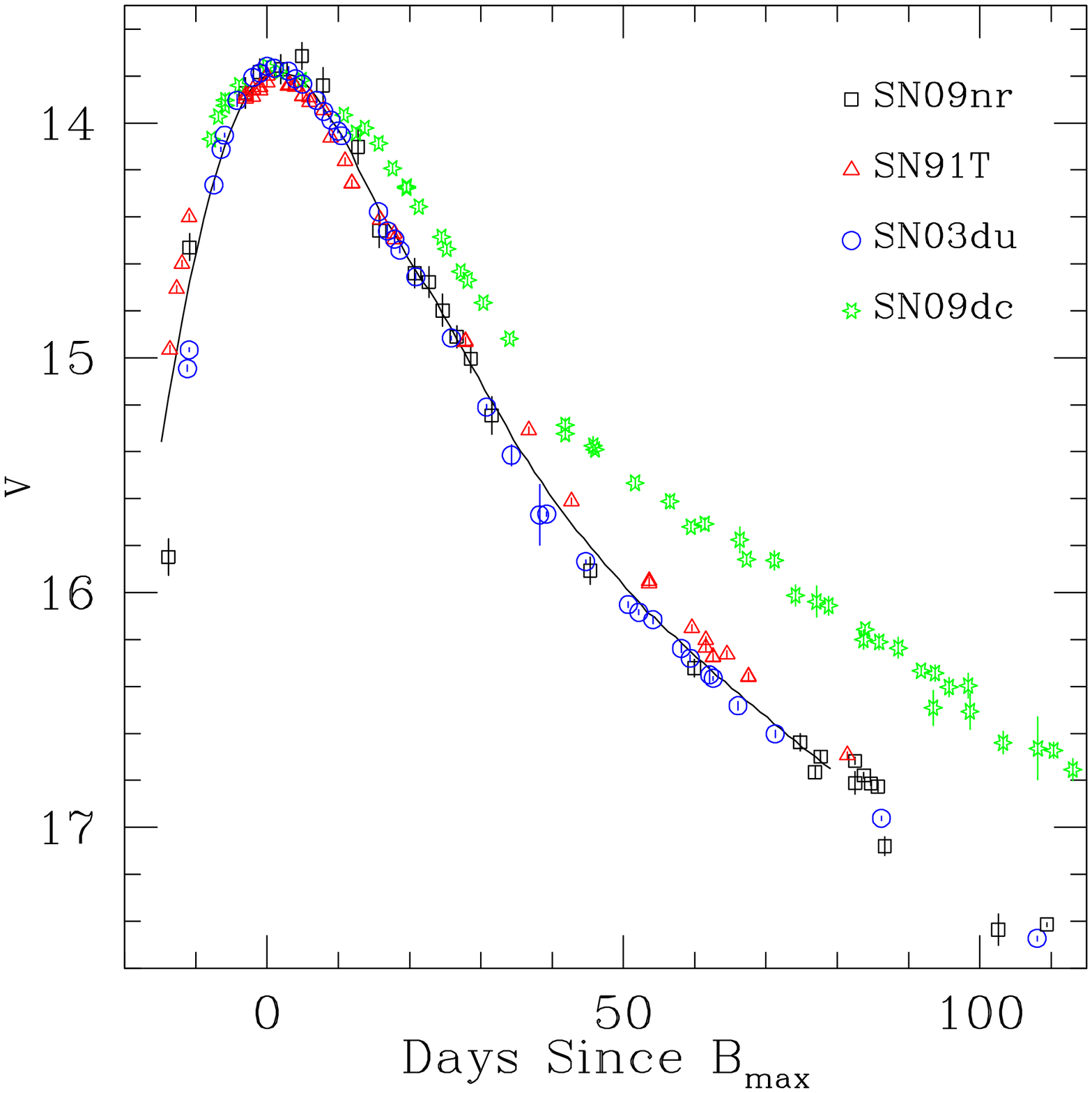}
\end{center}
\caption{Comparison of V-band light curve of SN~2009nr with those of
  the over-luminous Type Ia SN 1991T~\citep{ref:Lira_1998}, the normal
  Type Ia SN 2003du~\citep{ref:Stanishev_2007}, and the possibly
  super-Chandrasekhar mass Type Ia SN
  2009dc~\citep{ref:Silverman_2010}. The black solid line is the SN~Ia light curve
  template fit to the ASAS data points only. The light curves of the 
  comparison SNe are shifted so that their peaks
  approximately match that of SN~2009nr.}
\label{fig:lc_comp_v}
\begin{center}
\plotone{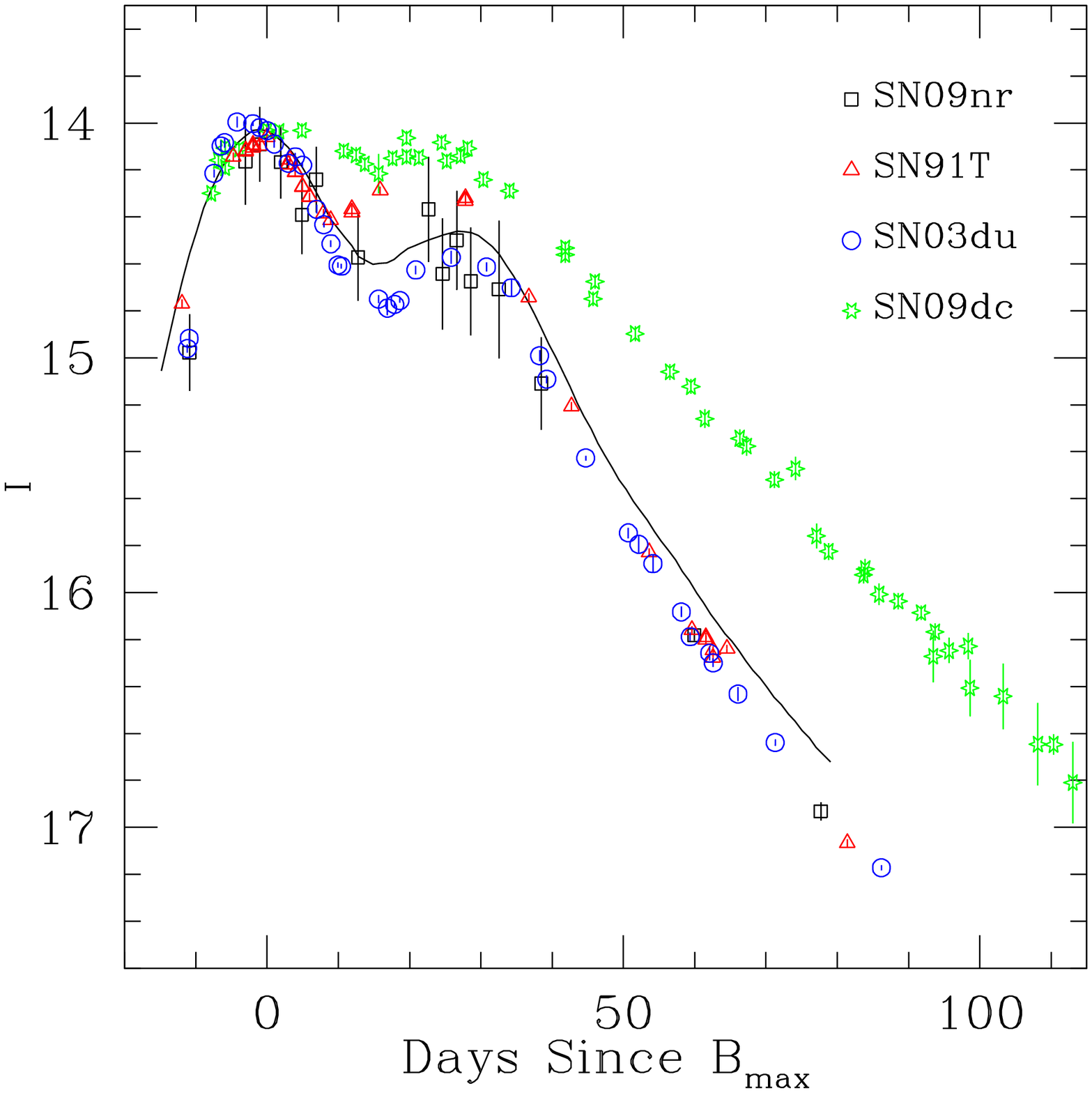}
\end{center}
\caption{Same as Figure~\ref{fig:lc_comp_v}, but for I-band.}
\label{fig:lc_comp_i}
\end{figure}

\subsection{Light Curve}
\label{sec:lightcurve}

\begin{figure*}[!t]
\begin{center}
{\includegraphics[scale=0.62]{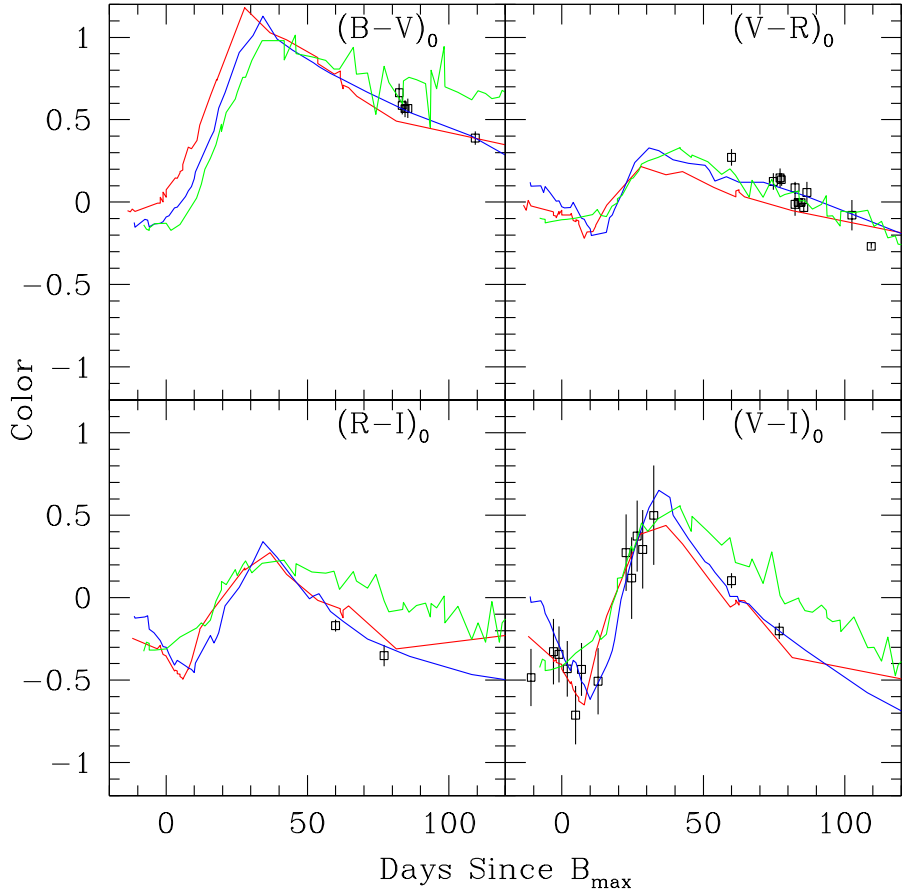}}
\end{center}
\caption{Color evolution of SN~2009nr (black data points) compared
  to those of the over-luminous Type Ia SN
  1991T~\citep[red;][]{ref:Lira_1998}, the normal Type Ia SN
  2003du~\citep[blue;][]{ref:Stanishev_2007}, and the possibly
  super-Chandrasekhar Type Ia SN
  2009dc~\citep[green;][]{ref:Silverman_2010}. The
  horizontal axis is same as in Figure~\ref{fig:lightcurve}. The continuous lines
  are connected color evolution data points of the respective SNe.}
\label{fig:colorcurve}
\end{figure*}

\begin{figure}[!ht]
\begin{center}
\plotone{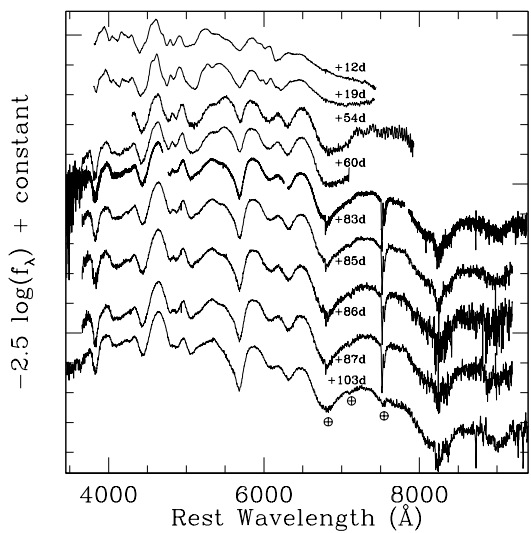}
\end{center}
\caption{Spectral evolution of SN~2009nr. Epochs are in days since the
  date of maximum brightness in the B-band. The vertical displacements are arbitrary. The telluric lines
  at 6800~\AA, 7200~\AA, and 7600~\AA~are labeled.}
\label{fig:spectra}
\end{figure}

\begin{figure}[!ht]
\begin{center}
\plotone{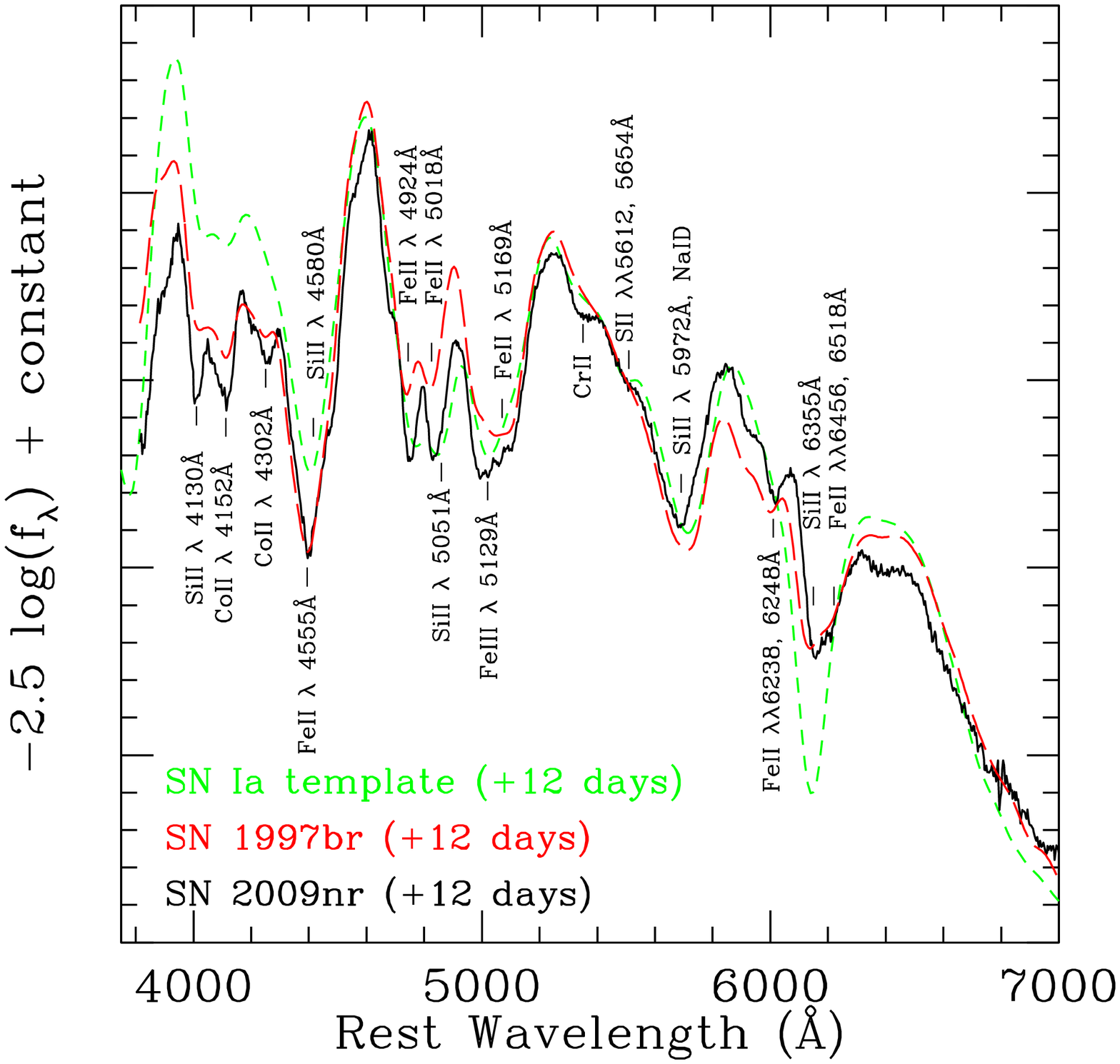}
\end{center}
\caption{Comparison of the earliest SN~2009nr spectrum (black continuous line) 
and the spectrum of SN~1991T like supernova 
SN~1997br~\citep[red long-dashed line, dereddened by $E(B-V)=0.35$ mag,][]{ref:Li_1999}, 
both collected 12 days after the date of maximum brightness in the B-band, with 
a normal SN~Ia spectrum template for the same post-maximum 
epoch~\citep[green short-dashed line;][]{ref:Hsiao_2007}. 
All the spectra are corrected to the rest-frame and shifted so that their flux 
approximately match at 5500~\AA.}
\label{fig:spec_type}
\begin{center}
\plotone{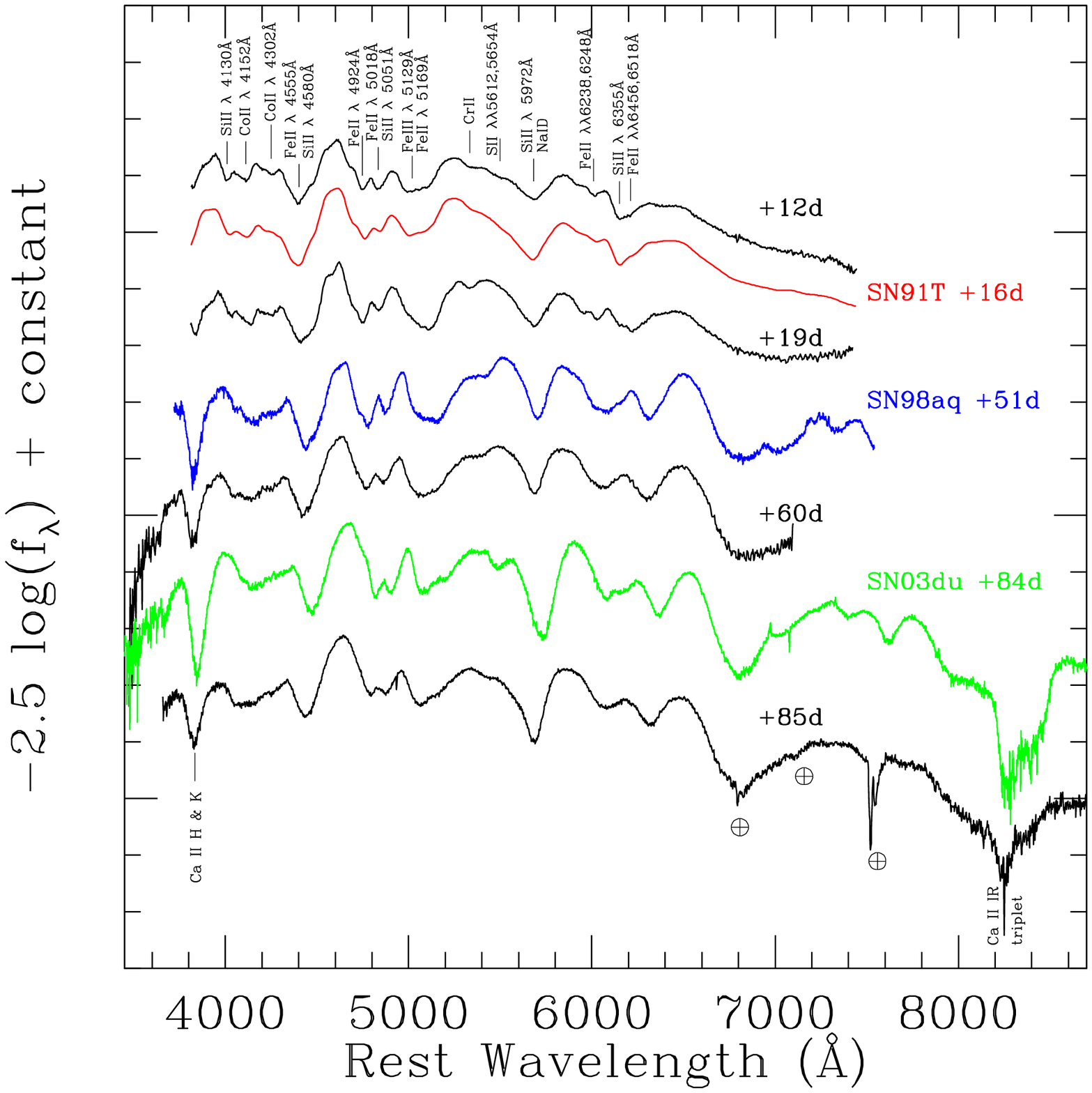}
\end{center}
\caption{Spectral evolution of SN~2009nr as compared to
  spectroscopically similar SNe at the relevant phase as determined by
  SNID~\citep{ref:Blondin_2007}. Important features are labeled along
  with the telluric lines (6800~\AA, 7200~\AA, and 7600~\AA).The data 
  for SN~1991T, SN~1998aq, and SN~2003du are from   \cite{ref:Mazzali_1995}, 
  \cite{ref:Branch_2003}, and \cite{ref:Stanishev_2007}, respectively. All 
  the spectra are corrected to the rest-frame using the redshifts ($z_{helio}$) 
  of the hosts. The SN~1991T spectra is dereddened by $E(B-V)=0.13$ mag~\citep{ref:Li_1999, ref:Saha_2001a}.
  Host extinction for SN~1998aq~\citep{ref:Reindl_2005}, SN~2003du~\citep{ref:Stanishev_2007}
  and SN~2009nr (Section~\ref{sec:lightcurve}) are consistent with zero. }
\label{fig:spec_comp}
\end{figure}

\begin{figure*}[!t]
\begin{center}
{\includegraphics[angle=0,scale=0.5]{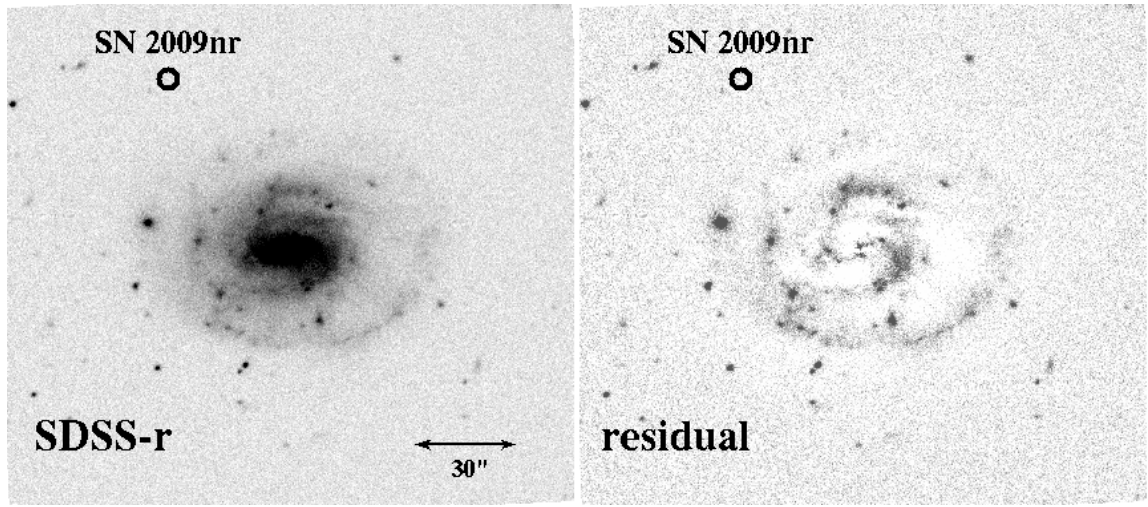}}
\end{center}
\caption{SDSS r-band image of the SN~2009nr host galaxy UGC~8255
  (left) and the residuals after subtracting a simple
  GALFIT~\citep{ref:Peng_2002} profile using a bulge/disk
  decomposition (Sersic and exponential disk profiles). The location
  of SN~2009nr is marked with a circle.}
\label{fig:galfit}
\end{figure*}

Figure \ref{fig:lightcurve} shows the V and I band light curves of
SN~2009nr. A SN~Ia light curve template fit to the ASAS V and I band
data was performed using the template fitting method described
in~\cite{ref:Prieto_2006}. The results of the fit are presented in
Table~\ref{table:fit_results}.

\begin{table}[h]
\begin{center}
\caption{Results of Light Curve Fitting}
\label{table:fit_results}
\begin{tabular}{lcl}
\hline 
\hline
\\
\multicolumn{1}{c}{Parameter} &
\multicolumn{1}{c}{Result} &
\multicolumn{1}{l}{Note}
\\
\hline
\hline
\\
${\chi^2_\nu}$ $(\nu=25)$ & $1.1$ & Chi-square of fit\\
$t_{max}(B)$ & $2455193.2\pm0.3$ & HJD of B-band maximum\\
$\Delta m_{15}(B)$ & $0.93\pm0.02$ mag & B-band decline rate\\
$E(B-V)_{host}$ & $0.00\pm0.01$ mag & Host color excess\\
$\mu$ & $33.27\pm0.15$ mag & Distance modulus
\\
\hline
\hline
\end{tabular}
\end{center}
\end{table}

The derived epoch of B-band peak magnitude, $t_{max}(B)$, implies that
the earliest ASAS measurement was obtained 14 days
before B-band maximum. The B-band magnitude decline in the
first 15 days after maximum light ($\Delta m_{15}(B)\simeq0.93$~mag)
confirms that SN~2009nr is more luminous than normal Type Ia SNe,
which usually have $\Delta
m_{15}(B)=1.1$~mag~\citep{ref:Phillips_1993}. The host galaxy
reddening, $E(B-V)_{host}$, estimate is consistent with zero, as
expected given the distance of the SN from the host (13.0 kpc,
$\sim4.3$ disk scale lengths).

The light curve fits imply peak apparent
magnitudes of $m_V=13.8$ mag and $m_I=14.0$ mag. This indicates a peak
absolute magnitude of $M_V\simeq-19.6$ mag and a distance modulus of
$\mu\simeq33.3$ mag ($d=45.1$ Mpc), based on the derived host galaxy
reddening of zero, Galactic foreground extinction of $A_V\simeq0.1$
mag, and Hubble constant H$_{0}=72$ km/s/Mpc. This is brighter than most
normal Type Ia SNe, and is consistent with that of over-luminous
ones~\citep[e.g.,][]{ref:Benetti_2005}.

Figures \ref{fig:lc_comp_v}, \ref{fig:lc_comp_i}, and
\ref{fig:colorcurve} compare the V and I band light curves and the
color evolution of SN~2009nr to those of the over luminous Type Ia SN
1991T~\citep[$M_V\simeq-19.6$;][]{ref:Filippenko_1992}, the normal
Type Ia SN 2003du~\citep[$M_V=-19.2$;][]{ref:Stanishev_2007}, and the
possibly super-Chandrasekhar Type Ia SN
2009dc~\citep[$M_V=-19.8$;][]{ref:Silverman_2010}. The light
curves are corrected to the CMB rest-frame and K-corrections are 
applied. The colors are corrected for Galactic foreground 
extinction, host galaxy reddening, and K-corrections. We only used the Bessell filters for 
K-corrections and did not apply S-corrections.

The SN~2009nr light curve is consistent with SN~1991T. It is clearly
different from SN~2009dc, which has a much slower decline
rate~\citep[$\Delta
  m_{15}(B)=0.72\pm0.03$~mag;][]{ref:Silverman_2010}, and it has a
very similar decline rate to SN 1991T~\citep[$\Delta
  m_{15}(B)=0.95\pm0.05$~mag;][]{ref:Phillips_1992}. The $B-V$ color
of SN~2009nr is consistent with the other SNe, while the $V-R$ color
is redder than the others between $\sim50$ to $\sim100$ days but becomes 
bluer around $\sim100$ days. The $R-I$ color is slightly bluer than the other
SNe between $\sim50$ to $\sim100$ days while the $V-I$ color is consistent
with the others.

A simple fit to the two earliest V-band data points (in flux scale) with 
assumed light curve rise going as $(t-t_{exp})^2$~\citep{ref:Arnett_1982,ref:Hayden_2010} 
implies $t_{exp}\simeq17.7$ days before B-band maximum. Since our light 
curve fits show that the V-band peak occurred at $t\simeq2$ days after 
B-band maximum, this indicates a rise time of $t_r=19.7$ days. Using 
the equation in Section~6.6 of~\cite{ref:Hayden_2010} with peak 
$M_V=-19.6$ mag and $t_r=19.7$ days gives a total $^{56}$Ni yield of 
$\sim0.9$~M$_{\odot}$, which is consistent with similar estimates for bright 
and slow SNe~Ia~\citep[e.g.,][]{ref:Stritzinger_2006,ref:Hayden_2010}.

\subsection{Spectra}
\label{sec:spectra}

The 9 post-maximum optical spectra of SN~2009nr span +12 to +103 days
relative to the epoch of B-band maximum light. Figure~\ref{fig:spectra} 
shows the temporal evolution of SN~2009nr spectra. Figure~\ref{fig:spec_type} 
compares the earliest SN~2009nr spectrum with a SN~1991T-like supernova and a 
normal SN~Ia template. 
Figure~\ref{fig:spec_comp}  compares the  SN~2009nr spectra with other SNe that are
determined to be spectroscopically similar at the relevant phase by
SNID~\citep{ref:Blondin_2007}. Important spectral features have been identified 
using \cite{ref:Li_1999}, \cite{ref:Pastorello_2007}, \cite{ref:Branch_2008}, and
\cite{ref:Wang_2009}.

SNID identifies the earliest SN~2009nr spectra collected 12 days after 
the date of maximum brightness in the B-band as spectroscopically most 
similar to a SN~1991T spectrum collected 16 days after maximum, while 
two SN~1991T-like SNe, SN~1997br and SN~1999aa (spectra collected 
12 and 15 days after peak, respectively), are identified as the second and 
third most similar spectra. In Figure~\ref{fig:spec_type}, we compare the 
earliest SN~2009nr spectrum and the spectrum of SN~1997br~\citep{ref:Li_1999} 
since both were collected 12 days after maximum. We also show a normal SN~Ia 
template for the same post-maximum epoch~\citep{ref:Hsiao_2007}. 
While many features of the earliest SN~2009nr spectrum are present in both 
the comparison spectra in Figure~\ref{fig:spec_type}, we specifically note 
some important features that are very similar in the SN~1997br and SN~2009nr 
spectra and differs from the normal SN~Ia template.

The \ion{Co}{2}~$\lambda$~4302\AA~and \ion{Fe}{2}~$\lambda\lambda$~6238, 
6248\AA~lines are clearly present in both the SNe spectra but in the 
normal SN~Ia template they are completely absent.
Also, the \ion{Si}{2}~$\lambda$~6355\AA~line of the two SNe spectrum, which 
is a characteristic feature of Type Ia SNe, are more consistent with each other 
than the normal SN~Ia template. As noted by \citet{ref:Li_1999} in case of 
SN~1997br, the red wing of the \ion{Si}{2}~$\lambda$~6355\AA~line of the 
SN~2009nr spectrum is also contaminated by the \ion{Fe}{2}~$\lambda\lambda$~6456, 
6518\AA~line, unlike the normal SN~Ia template where the \ion{Si}{2} line is 
much stronger. The \ion{Co}{2}~$\lambda$~4152\AA, \ion{Fe}{2}~$\lambda$~4555\AA,
\ion{Fe}{2}~$\lambda$~4924\AA, and \ion{Fe}{2}~$\lambda$~5018\AA~lines 
of the two SNe spectra appear stronger while the \ion{Si}{2}~$\lambda$~4580\AA,
\ion{Si}{2}~$\lambda$~5051\AA, and \ion{S}{2}~$\lambda\lambda$~5612, 5654\AA~lines 
are more prominent in the normal SN~Ia template.

Based on the SNID comparison and the better match of spectral features in 
the earliest spectra (as well as the B-band decline rate and peak magnitude) 
we conclude that SN~2009nr is most likely a SN~1991T-like object.

Although at early times SN~2009nr is spectroscopically most similar to
the over-luminous Type Ia SN~1991T, at later times it becomes more
similar to the normal Type Ia SN~1998aq and SN~2003du. This is 
consistent with the observed characteristics of 1991T-like SNe
(not very strong \ion{Si}{2}, different ionization state of some dominant
features because of higher temperature etc.) that are strongest before and
around maximum light, but later on dilute to become similar to normal
Type Ia SNe. The earliest spectrum at $t=+12$ days shows
singly-ionized lines of intermediate mass elements (Si, S, and Na), as
well as various \ion{Fe}{2} lines. The \ion{Si}{2}~$\lambda$~6355\AA~line is
clearly visible at $t=+12$ days, and is identifiable at $t=+19$ days
as well, but not at $t=+54$ days. The \ion{Cr}{2} line is visible at $t=+12$
days, and becomes more prominent at $t=+19$ days, but is no longer
identifiable in the iron group element dominated latter spectra.

We examined the blueshifted \ion{Si}{2}~$\lambda$~6355\AA~and \ion{Fe}{2}
$\lambda$~4555\AA~absorption features in the earliest spectra ($t=+12$
days) to determine the photospheric expansion velocity of
SN~2009nr. The \ion{Si}{2} line is observed at redshift corrected
$\lambda$~6150\AA~implying an expansion velocity of $\sim9700$ km/s,
and the \ion{Fe}{2} line is observed at redshift corrected
$\lambda$~4400\AA~implying an expansion velocity of $\sim10000$ km/s,
which are consistent with the velocities of normal and over-luminous
Type Ia SNe~\citep[e.g.,][]{ref:Benetti_2005} at similar epochs.

\subsection{Host Galaxy Properties}
\label{sec:galaxy}

Table~\ref{table:host_properties} summarizes the basic parameters of the host galaxy of
SN~2009nr, UGC~8255. We derived total magnitudes of UGC~8255
from SDSS in the $ugriz$ bands using
Sextractor~\citep{ref:Bertin_1996}. The total SDSS $gr$ magnitudes
were used to infer the absolute magnitude in $B$, correcting for
Galactic extinction \citep{ref:Schlegel_1998}, and
K-corrections~\citep{ref:Blanton_2007}, and using the distance modulus
derived from the light curve fits (see
Table~\ref{table:fit_results}). The derived absolute magnitude in $B$
and the rest-frame $B-V$ color of UGC~8255 are shown in
Table~\ref{table:host_properties}.

We used the total magnitudes of UGC~8255 in the SDSS bands and GALEX
UV photometry (FUV and NUV), estimated from aperture photometry, to
fit stellar population synthesis models (SPS) with the FAST
code~\citep{ref:Kriek_2009}. The results for the total stellar mass,
age, and star-formation rate derived from the SPS fits with FAST are
also presented in Table~\ref{table:host_properties}. Within FAST we used the
\cite{ref:Bruzual_2003} SPS models, a Salpeter IMF, and assumed solar
metallicity based on abundance estimates we discuss shortly.

Figure \ref{fig:galfit} shows the SDSS r-band image of the SN~2009nr
host galaxy UGC~8255, and the residual after subtracting a
GALFIT~\citep{ref:Peng_2002} model that includes an exponential disk
and a Sersic profile for the bulge. The results are also presented in
Table~\ref{table:host_properties}.

We took a spectrum of the host galaxy using IMACS on the Baade 6.5-m
Telescope at the Las Campanas Observatory, running the slit across the
supernova and the center of the host. Figure \ref{fig:host_oh} shows
the radial metallicity dependence of the SN~2009nr host galaxy,
UGC~8255, along the slit. Oxygen abundances of the nucleus and 6
off-center H~II regions were estimated using the [N~II]/H$\alpha$
calibration described in~\cite{ref:Denicolo_2002}. The linear fit to the
Oxygen abundance measurements has a gradient of $\sim0.06$ dex/kpc,
which is consistent with other observations for similar galaxy
type~\citep{ref:Zaritsky_1994}.

\section{Discussion}
\label{sec:discussion}

\subsection{SN~2009nr Properties}
\label{sec:sn09nr}

\begin{figure}[!t]
\begin{center}
\plotone{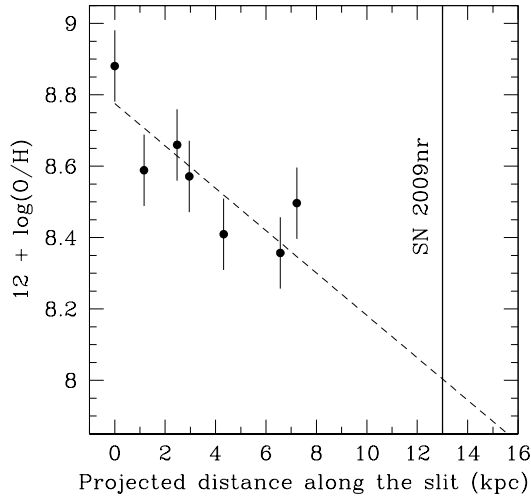}
\end{center}
\caption{Radial oxygen abundance profile of the SN~2009nr host galaxy
  UGC~8255 along a slit going through the SN and the center of the
  host (based on 6 off-center H~II regions). The Oxygen abundance is
  calculated using the [N~II]/H$\alpha$ ratio. The dashed line shows a
  linear fit to the Oxygen abundance measurements and has a gradient
  of $\sim0.06$ dex/kpc. The solid line marks the position of the SN.}
\label{fig:host_oh}
\end{figure}

SN~2009nr has a projected distance of 13.0 kpc or $\sim4.3$ disk scale
lengths from the nucleus of its metal rich star-forming Scd host
galaxy UGC~8255. This is a lower limit on its physical distance, since
we do not know if the object is located on the same plane as the
disk. The derived physical parameters of the galaxy, the radial
metallicity gradient, and the projected distance of the SN from its
host nucleus all indicate that SN~2009nr is located in the halo of
UGC~8255 rather than in an extended disk. Given that the metallicity
of Milky Way halo stars is $\sim1$ dex lower than the disk
stars~\citep[e.g.,][]{ref:Ivezic_2008}, and the radial metallicity
gradient of UGC~8255 (Figure \ref{fig:host_oh}), it is apparent that
SN~2009nr exploded in a low metallicity region.

\begin{table}[!b]
\begin{center}
\caption{SN~2009nr Host Properties}
\label{table:host_properties}
\begin{tabular}{lcl}
\hline 
\hline
\\
\multicolumn{1}{c}{Property} &
\multicolumn{1}{c}{Value} &
\multicolumn{1}{c}{Note}
\\
\hline
\hline
\\
Name & UGC~8255 & \\
R.A. (J2000.0) & 13$^h$10$^m$56.5\fs95 & \cite{ref:SDSS_2007} \\
Dec. (J2000.0) & +11\arcdeg28\arcmin38\farcs5 & \cite{ref:SDSS_2007} \\
$z_{CMB}$ & 0.0122 & \cite{ref:Fixsen_1996}\\
$z_{helio}$ & 0.0113 & \cite{ref:RC3_1991}\\
Galaxy type & Scd & \cite{ref:RC3_1991}\\
$M_B$ & $-19.05$ mag & Absolute B-band magnitude\\
Color $(B-V)_{0}$ & 0.52 mag & Corrected to the rest frame\\
log$(M_{\star}/M_\odot)$ & 9.60 & [9.51, 9.71]\tablenotemark{a}\\
log(SFR/$(M_\odot/yr)$) & $-0.35$ & [$-0.42$, $-0.22$]\tablenotemark{a}\\
log(Age/yrs) & 9.14 & [8.98, 9.31]\tablenotemark{a}\\
12 + log(O/H) & 8.78 &  Central oxygen abundance\\
$R_s$ & $14\farcs3$ (3.0 kpc) & Exponential disk scale\\
$b/a$ & 0.77 & Axis ratio (exponential)\\
$PA$  & $86.1\arcdeg$ & Position angle (exponential)\\
Sersic Index & 4.24 & For Sersic profile\\
$b/a$ & $0.49$ & Axis ratio (Sersic)
\\
\hline
\hline
\end{tabular}
\end{center}
\tablenotetext{a}{$1\sigma$ range.}
\end{table}

The spectroscopic identification, B-band decline rate, and absolute
magnitude all indicate that SN~2009nr is a SN~1991T-like
event. Significantly more luminous than normal Type Ia SNe, SN~1991T
is thought to be the result of the delayed detonation of a CO WD, or
the double detonation of a WD initiated at the boundary layer between
the CO core and the He envelope~\citep{ref:Filippenko_1992}. More 
recently, it has been proposed that the SN~1991T-like SNe progenitor 
population may emerge from WD+MS systems~\citep{ref:Kasen_2004}, 
and that these events may not have any special physical properties 
except for the viewing angle of the observer~\citep{ref:Meng_2010}.

If the SN~2009nr progenitor formed in the central region of the host
and lived few$\times100$~Myr, then it would need to have a rather
implausible radial velocity of 120 km/s to travel 13.0 kpc in its
lifetime. If it traveled that far from the host at a much slower
speed, then the progenitor must be an old star. Alternatively, the 
progenitor formed in the metal poor environment of the halo, but then
it cannot be related to the young population located in the host
galaxy's central region. While we can not tell if the progenitor 
formed in the disk and moved out or it formed in the halo long ago, 
it would be old in either case. Therefore, we cannot make any statement
regarding the stellar population from which the progenitor emerged,
based on the recent star formation rate or environment of the host.

\subsection{Implication for Prompt Type Ia SNe}
\label{sec:prompt}

Studies of the host metallicities of nearby SNe
Ia~\citep{ref:Hamuy_2000,ref:Gallagher_2005} have claimed that the
primary factor regulating SN~Ia peak luminosities is the age of the
stellar population rather than metallicity. A two-component progenitor
distribution has been proposed, each associated with a distinct
evolutionary timescale~\citep[e.g.,][]{ref:Mannucci_2005,
  ref:Scannapieco_2005, ref:Neill_2006,ref:Maoz_2010}. The ``prompt''
component is thought to be correlated with the recent star formation
rate of the host, and the SN explodes $\sim100$ Myr after star
formation. This leads to the high SNe~Ia rates in actively star
forming galaxies (late type spirals and irregulars). The ``delayed''
component tracks the underlying stellar population, scales with
stellar mass, and explodes $\gtrsim 1$~Gyr after star formation, usually
in old, quiescent, elliptical galaxies. The delayed component can be 
much older, as noted in recent delay time distribution 
studies~\citep[e.g.,][]{ref:Maoz_2010a, ref:Horiuchi_2010}.

Luminous SNe~Ia are conventionally associated with the prompt
component, supposedly emerging from young stellar populations
\citep{ref:Hamuy_2000,ref:Cooper_2009,ref:Sullivan_2010},
and the brightest events explode mainly in star forming 
galaxies~\citep{ref:Hamuy_1996, ref:VanDenBergh_2005}. 
However, given its location, SN~2009nr has to
be coming from an old population, and is clearly unrelated to the
host's recent star formation history.  In fact, Type Ia SNe that are 
distant from their host nucleus are not rare. Figure 21 of 
\cite{ref:Hicken_2009} shows that a significant
fraction (roughly, a third) of SNe~Ia are more than 10 kpc away from
their star forming hosts. This is true both for SNe located in spirals
or irregular galaxies and similar to SN~2009nr (the brightest, bluest,
slowest ones; the blue points in that figure), and also for those in
the broader host population with more normal luminosities (the green
points in that figure). 

\cite{ref:Prieto_2008} discusses some extreme and interesting SN-host 
pairs, such as the the metal-poor ($1/4$ solar) host of SN 2007bk, 
where the 1991T-like over-luminous SN was found $\sim9$~kpc away 
from its dwarf host galaxy's center. \cite{ref:Badenes_2009} demonstrates 
that three of the four young SN~Ia remnants in the LMC are associated with old,
metal-poor stellar populations. For example, SNR~$0509-67.5$ is known to have 
been originated by an extremely bright Type Ia event but is located in a 
population with a mean age of 7.9~Gyr, far away from any sites of recent star 
formation. Moreover, three of the four claimed Super-Chandrasekhar SNe detected 
in recent years, all extremely bright Type Ia events, are located far away from 
their star forming hosts. Of these, 
SN~2003fg~\citep[$M_V=-20.0$;][]{ref:Howell_2006} is located 0.9~kpc away from 
its small low-mass star-forming host, 
SN~2006gz~\citep[$M_V=-19.2$;][]{ref:Hicken_2007} is located 14.4~kpc away from 
its Scd host, and SN~2009dc~\citep[$M_V=-19.8$;][]{ref:Silverman_2010} is 
located $\sim12$~kpc (5.7 Petrosian Radii) away from its S0 host. Besides,
SBS~$1150+599$A, a close binary system hosted by the planetary nebula 
PNG~$135.9+55.9$ located in the Galactic halo~\citep{ref:Tovmassian_2010}, 
is an excellent example of a possible Type Ia SN progenitor candidate 
in a very low metallicity environment of a star-forming host (Milky 
Way, in this case) that is unrelated to recent star formation.

Given the location of the progenitors and properties
of their hosts, it appears to us that the higher SN luminosities are
likely related to the low metallicity
local environments of these SNe progenitor systems rather than
just the age of the stellar population~\citep[see][for related discussion]{ref:Prieto_2008}.
This motivates us to rethink the conventional association of 
luminous SNe~Ia with the ``prompt'' component directly correlated with 
recent star formation.

\subsection{Utility of ASAS for Studying Nearby SNe}
\label{sec:asas}

\begin{figure}
\begin{center}
\plotone{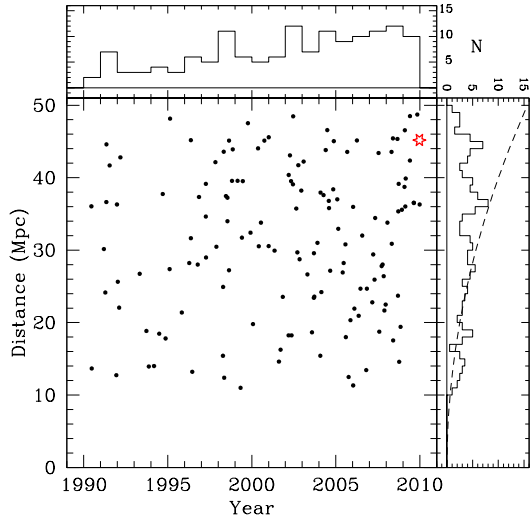}
\end{center}
\caption{History of Type Ia SNe discovery within d $\lesssim50$~Mpc
  over the past twenty years according to the Sternberg Astronomical
  Institute SN Catalogue~\citep{ref:Bertunov_2007}. SN 2009nr is shown 
  with a red starred symbol. The 
  histograms show projections on the respective axes. The dashed
  line in the histogram to the right indicates the $z=0$ extrapolation
  of the cosmic SN~Ia rate density~\citep{ref:Horiuchi_2010} for this
  local volume. (The distances were obtained assuming the Hubble flow. 
  Because of the importance of SN~Ia at the closest distances, two 
  objects apparently within 10~Mpc were examined separately and found 
  to have distance estimates in the literature beyond 10 Mpc, and were 
  corrected. See~\cite{ref:Horiuchi_2010} for further discussion on 
  nearby SN~Ia rates.)}
\label{fig:dist_yr}
\end{figure}

Clearly, the conventionally accepted understanding of luminous Ia SNe
being directly related to the recent star forming environment of
their hosts needs to be reconsidered. If many of the luminous Ia like
SNe are indeed physically separated from the central star forming
regions of their hosts, efforts to establish causal links between SNe
Ia and their host environments become more challenging. For example,
\cite{ref:Brandt_2010} discusses the host properties of SNe~Ia
discovered by SDSS, but SN~2009nr is an obvious example in which the
SN would be outside the SDSS fiber, and we cannot learn anything
about the  progenitor environment (age and metallicity) from the SDSS 
spectrum. Similarly, \cite{ref:Gallagher_2005}, \cite{ref:Howell_2009}, 
and \cite{ref:Neill_2009} discuss
relationships between SNe~Ia and their star-forming host metallicity, in the
context of ``global'' measurements of oxygen abundance based on the
mass of the galaxy or luminosity-weighted oxygen abundances. These may 
not be very good approximations for SNe located in parts of their hosts 
that have significantly different metallicity and star forming history 
than what is inferred from such global estimates~\citep[e.g.,][]{ref:Badenes_2009}.

Although there is general agreement that SNe~Ia are thermonuclear
explosions of WDs, whether it involves only one degenerate object
stripping mass off its hydrogen-rich binary companion until it
explodes upon reaching Chandrasekhar Mass, or two WDs merging into a
super-Chandrasekhar mass object which then explodes, is less
certain. Conceivably, either of these mechanisms can be involved,
although it has been proposed that only one of them might 
dominate~\citep{ref:DiStefano_2010,ref:DiStefano_2010a,ref:Gilfanov_2010,ref:Hayden_2010a}.

\cite{ref:Pritchet_2008} demonstrates that SNe~Ia explosions are about
1\% of the stellar death rate independent of star formation history,
and proposes that some progenitor scenario other than the
single-degenerate channel alone must be invoked to explain SNe 
Ia. \cite{ref:Ruiter_2009} finds strong indication that the double
degenerates form the dominant channel generating Type Ia SNe in spiral
galaxies. \cite{ref:Mennekens_2010} concludes that while the double
degenerate delay time distribution, possibly combined with the single
degenerate one, agrees with observation, the single degenerate
scenario alone cannot reproduce the observed distribution. Detection
of a greater number of, if not all, nearby ($d\lesssim50$ Mpc) SNe,
for which it is possible to obtain very late time spectra, can help
refine our current interpretation of what physical phenomenon may be
contributing to the prompt and delayed SN~Ia
population~\citep[e.g.,][]{ref:Leonard_2007}.

While surveys targeting large and bright galaxies lead to a higher SN
detection rate per observed galaxy, the low metallicity, low
luminosity galaxies have hosted some of the most interesting SNe of
all types (for example, SN~1999aw~\citep{ref:Strolger_2002},
SN~2005cg~\citep{ref:Quimby_2006}, SN~2005gj~\citep{ref:Prieto_2007},
SN~2007bk~\citep{ref:Prieto_2008},
SDWFS-MT-1~\citep{ref:Kozlowski_2010} etc.). To identify and analyse
the most unusual and potentially most informative SNe, we need
galaxy-impartial surveys. High cadence all sky surveys conducted using
small telescopes such as ASAS give us the opportunity to build a
galaxy independent SN sample, even if only for the very local
universe. As illustrated in Figure~\ref{fig:dist_yr}, although the SN
Ia detection rate has risen significantly in recent years,
incompleteness continues to be a very real concern even in the
immediate neighborhood ($d\lesssim50$ Mpc).

The observation of SN~2009nr at $d\simeq45$ Mpc two weeks prior to peak
luminosity demonstrates the potential capability of ASAS, and similar
high-cadence all sky surveys conducted using small telescopes, to
meaningfully address this incompleteness issue. Moreover, early
detection and well sampled photometric data around the peak will
enable us to accurately reconstruct the light curves and determine the
peak luminosity, which can potentially address the theoretical
uncertainty over SN~Ia progenitor mass in the double degenerate
scenario. Discovery of most nearby SNe~Ia for which we may obtain very
late time spectroscopic data, which is not possible to do for distant
SNe, can also help us better understand the progenitor systems.

\section{Conclusions}
\label{sec:conclusion}

SN~2009nr is a SN~1991T-like supernova that is more luminous and has a slower initial decline rate than normal Type Ia SNe. Located $\sim4.3$ disk scale lengths away from the nucleus of its star-forming host, it either formed in the halo or wandered out over a long time. Evidently, it is not associated with the young stellar population and central star forming environment of its host.

In fact, many bright and slow SNe~Ia occurring in star-forming hosts are not associated with young stellar populations, and thus should not be considered a part of the prompt Ia population by default. Type Ia SNe that are located far from their host galaxy nucleus probably have no association with the recent star forming history of that galaxy. This may affect attempts to explore causal connections between SNe~Ia and their host properties.

Scientifically interesting SNe, of both Type I and II, have often been discovered in low metallicity, low luminosity galaxies that are usually not targeted in SN surveys. Galaxy-impartial high cadence all sky searches conducted using small telescopes such as ASAS can produce a galaxy independent SN sample and lead to pre-maximum detection of many local ($d\lesssim50$ Mpc) SNe.

\acknowledgments

We thank the referee, Xiangcun Meng, and Andrew Drake for helpful comments, C.~S.~Kochanek for helpful comments and discussions, and M. Kriek for making her SPS code publicly available. We are grateful to the staffs of the Las Campanas Observatory, the Apache Point Observatory, the MDM Observatory, and the Winer Observatory for their excellent support. This research has made use of NED, which is operated by the JPL and Caltech, under contract with NASA and the HEASARC Online Service, provided by NASA's GSFC. RK and KZS are supported in part by NSF grant AST-0707982. JLP acknowledges support from NASA through Hubble Fellowship grant HF-51261.01-A awarded by the STScI, which is operated by AURA, Inc. for NASA, under contract NAS 5-26555. GP and BP are supported by the Polish MNiSW grant N203 007 31/1328. KZS and DMS are supported in part by NSF grant AST-0908816. JFB is supported by NSF CAREER grant PHY-0547102.

\bibliographystyle{apj}

\bibliography{bibliography}

\end{document}